\documentclass[twocolumn,aps,prl,superscriptaddress]{revtex4}
\usepackage{amssymb}
\usepackage{amsmath,bm}
\usepackage{graphicx}
\usepackage[normalem]{ulem}
\usepackage{multirow}
\usepackage[colorlinks,linkcolor=blue,urlcolor=blue,citecolor=blue]{hyperref}
\usepackage{lipsum}
\usepackage[usenames, dvipsnames]{xcolor}
\usepackage{tensor}
\usepackage{isotope}
\usepackage{amsmath}
\allowdisplaybreaks[4]
 
\renewcommand{\sout}{\bgroup \color{red} \ULdepth=-.5ex \ULset}

\begin{document}

\title{Unveiling  the  dynamics of nucleosynthesis in relativistic heavy-ion collisions}

\author{Kai-Jia Sun}
\email{kjsun@tamu.edu}
\affiliation{Cyclotron Institute and Department of Physics and Astronomy, Texas A\&M University, College Station, Texas 77843, USA}

\author{Rui Wang}
\email{wangrui@sinap.ac.cn}
\thanks{\\Kai-Jia Sun and Rui Wang contributed equally to this work and should be considered as co-first authors.}
\affiliation{Key Laboratory of Nuclear Physics and Ion-beam Application~(MOE), Institute of Modern Physics, Fudan University, Shanghai $200433$, China}

\author{Che Ming Ko}
\email{ko@comp.tamu.edu}
\affiliation{Cyclotron Institute and Department of Physics and Astronomy, Texas A\&M University, College Station, Texas 77843, USA}

\author{Yu-Gang Ma}
\email{mayugang@fudan.edu.cn}
\affiliation{Key Laboratory of Nuclear Physics and Ion-beam Application~(MOE), Institute of Modern Physics, Fudan University, Shanghai $200433$, China}
\affiliation{Shanghai Research Center for Theoretical Nuclear Physics, NSFC and Fudan University, Shanghai 200438, China}

\author{Chun Shen}
\email{chunshen@wayne.edu}
\affiliation{Department of Physics and Astronomy, Wayne State University, Detroit, MI 48201, USA}
\affiliation{RIKEN BNL Research Center, Brookhaven National Laboratory, Upton, NY 11973, USA}

\date{\today}
\begin{abstract}
Like nucleosynthesis  during the early universe, light nuclei are also produced in relativistic heavy-ion collisions. Although the deuteron ($d$)  yields in these collisions can be well described by the statistical hadronization model (SHM), which assumes that particle yields are fixed at a common chemical freezeout near the phase boundary between the quark-gluon plasma and the hadron gas, the  recently measured  triton ($^3\text{H}$) yields   in Au+Au collisions at $\sqrt{s_{NN}}=7.7-200$ GeV are overestimated systematically by  this model.  Here, we develop a comprehensive kinetic approach to study the effects of hadronic re-scatterings, such as  $\pi NN\leftrightarrow\pi d$ and $\pi NNN\leftrightarrow\pi ^3\text{H}~(^3\text{He})$, on $d$, $^3\text{H}$, and $^3\text{He}$ production in these collisions. We find that these reactions have little effects on the deuteron yield but reduce the $^3\text{H}$ and $^3\text{He}$ yields by about a factor of 1.8 from their initial values given by the SHM. This finding helps resolve the overestimation of triton production in the SHM and provides the evidence for hadronic re-scattering   effects  on nucleosynthesis in relativistic heavy-ion collisions.
\end{abstract}

\maketitle
\emph{Introduction.}{\bf ---}
In the big bang nucleosynthesis~\cite{Olive:1999ij}, light nuclei such as deuterons~($d$), $^3\text{H}$, $^3\text{He}$, and $^4\text{He}$ are formed through a sequence of two-body reactions like $pn\leftrightarrow \gamma d$ at temperatures less than  1 MeV. These  nuclei and their anti-partners can also be produced in relativistic heavy-ion collisions~\cite{STARSc328,STARNt473,ALICE:2015rey,STARNtP16}, but at much higher temperatures  of  $\sim 100$ MeV.  Despite of extensive works over the years, their production mechanisms, especially  regarding  the role of hadronic re-scatterings, in these collisions are still  not understood  ~\cite{CsePR131,CJHPR760,Ono:2019jxm,Braun-Munzinger:2018hat,Schukraft:2017nbn,Xu:2018jff,Rais:2022gfg}.

In  relativistic heavy-ion collisions~\cite{Heinz:2013wva}, the produced hot dense matter undergoes various stages of evolution including the  pre-equilibrium  dynamics, partonic expansion, hadronization of the partonic matter, which can be modelled by the statistical hadronization model (SHM)~\cite{AndNt561}, and subsequent hadronic re-scatterings and resonance decays until the kinetic freeze-out.  In SHM, light nuclei, like ordinary hadrons, are assumed to be thermally produced from the quark-gluon plasma (QGP) and remain intact during later hadronic evolution.  However, considering the large cross section of $\sim 100~$mb~\cite{PDG2020} for deuteron dissociation by a pion and its inverse  reaction, i.e., $\pi d \leftrightarrow \pi NN$, deuterons would be disintegrated and regenerated continuously during the hadronic matter expansion. The recent extensions  of the SHM to include the hadronic effects using the Saha equation~\cite{Vovchenko:2019aoz}  or the rate equation~\cite{Neidig:2021bal} have demonstrated that the $d$ and $^3\text{H}$ ($^3\text{He}$) yields remain essentially unchanged from the chemical freezeout near the phase boundary to the kinetic freezeout.  A similar conclusion on (anti-)deuteron production is obtained in a more sophisticated kinetic or transport approach~\cite{DanNPA533,OhPRC76,OhPRC80,OliPRC99}, in which the reactions  $\pi d \leftrightarrow \pi NN$ are explicitly included during the hadronic expansion ~\cite{OliPRC99}. These theoretical studies thus suggest that neglecting the hadronic effects, as assumed in the SHM, is a good approximation.  Although the SHM describes quite well deuteron production in relativistic heavy-ion collisions~\cite{Braun-Munzinger:2018hat}, it has been recently found to systematically overestimate the triton and helium-3 yields measured in central Au+Au  and Pb+Pb  collisions at $\sqrt{s_{NN}}=6.3-200$ GeV~\cite{Zhang:2020ewj,NA49:2016qvu} by about a factor of 2  even after  the inclusion of feed-down contributions from unstable nuclei~\cite{Vovchenko:2020dmv}.

To understand the origin of the overestimation, we  re-examine the effects of hadronic re-scatterings on  $^3\text{H}$ ($^3\text{He}$) production, which have been neglected in the SHM, through more realistic dynamical approaches like the kinetic equations. Due  to the difficulty in treating many-body scatterings, no attempt has yet been made to include  in the kinetic approach the more complicated four-body reactions $\pi NNN\leftrightarrow \pi^3\text{H}$ ($\pi^3\text{He}$) for $^3\text{H}$ ($^3\text{He}$) production.  Also, previous studies based on the kinetic  approach have all neglected the finite sizes of light nuclei, which have been shown in the nucleon coalescence model for nuclei production to significantly suppress the $d$ and $^3$He yields in collisions of small systems like p+p collisions~\cite{SKJPLB792,Bellini:2020cbj}. In the present study, we develop a more realistic kinetic approach by including the finite sizes of nuclei to study the effects of hadronic re-scatterings on $d$, $^3\text{H}$, and $^3\text{He}$ production in relativistic heavy-ion collisions.  This approach is based on an extension of the relativistic kinetic equations, derived in Ref.~\cite{DanNPA533} from the real-time Green's functions~\cite{Danielewicz:1982kk,DANIELEWICZ1990154,Buss:2011mx}. The resulting nonlocal collision integrals are  evaluated by a stochastic method. With this approach, we find that the initial triton and helium-3 yields from the SHM are reduced by   a factor of 1.8 during the hadronic matter expansion in central Au+Au and Pb+Pb collisions at $\sqrt{s_{NN}}=20-5020$ GeV, and their final values are in good agreements with  measured values in experiments.

\begin{figure}[t]
  \centering 
   \includegraphics[width=6.1cm]{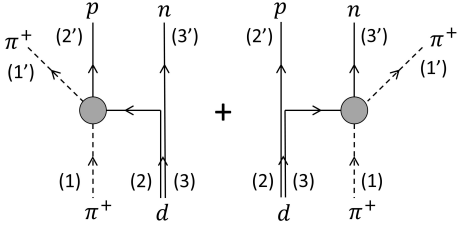} 
  \caption{  Diagrams for the reaction  $\pi^+d\leftrightarrow \pi^+np$ in the impulse approximation. The filled bubble indicates the intermediate states such as a $\Delta$ resonance. }
  \label{pic:piond}
\end{figure}

\emph{A  relativistic  kinetic approach.}{\bf ---}We illustrate our approach  by considering   the  specific channel $\pi^+d\to \pi^+np$ for deuteron dissociation. With the typical temperature ($T$) of the hadronic matter in high-energy nuclear collisions at  100-150 MeV,  the pion thermal wavelength  is  around  0.4-0.5 fm, which is much smaller than the deuteron diameter or size of about 4 fm.  A pion thus has a sufficiently large momentum to resolve the two constituent nucleons in the deuteron, resulting its scattering by one nucleon with the other nucleon being a spectator, as shown in Fig.~\ref{pic:piond}. This quasifree or impulse  approximation (IA) was previously used in studying deuteron dissociation in low-energy heavy-ion collisions~\cite{ChePR80, DanNPA533} and also  $J/\Psi$ dissociation by partons in relativistic heavy-ion collisions~\cite{GraPLB523,Du:2019xqq}. Under this approximation, the inverse reaction $\pi^+ np\to\pi^+ d$ can be viewed as a  two-step process, i.e., the scattering between a pion and a nucleon with the final-state nucleon sightly off the mass shell  and then subsequently coalescing with another nucleon to form a deuteron. 

Specifically, the production and dissociation  of deuterons  in  a pion dominated  hadronic matter can be similarly described by the relativistic kinetic equation (RKE), previously derived in Refs.~\cite{DanNPA533} for a pure nucleonic matter that involves the reactions $NNN\leftrightarrow Nd$, i.e.,
\begin{eqnarray}
\frac{\partial f_d}{\partial t}+\frac{  {\bf P}}{\  {E_d}}\cdot \frac{\partial f_d}{\partial {\bf R}}=-\mathcal{K}^{>}f_d + \mathcal{K}^{<}(1+f_d), 
 \label{eq:deu_kin}
\end{eqnarray}
where the deuteron distribution function $f_d({\bf R},{\bf P})$ is normalized as $g_d(2\pi)^{-3}\int \text{d}^3{\bf R}\text{d}^3{\bf P}f_d=N_d$ with $g_d=3$ and $N_d$ being the deuteron  spin degeneracy  factor and total yield, respectively.  On the l.h.s. of Eq.~(\ref{eq:deu_kin}), which denotes the drift term, we have treated the deuteron as a free particle by neglecting its mass change and mean-field potential. The two terms on the r.h.s. of Eq.~(\ref{eq:deu_kin}) describe the deuteron dissociation (${\cal K}^>$) and production (${\cal K}^<$) rates, respectively. Under the impulse approximation, the collision integral on the r.h.s. is given by~\cite{DanNPA533,DanPLB274} 
 \begin{eqnarray}
&&\frac{1}{2g_dE_d}\int \prod_{i=1'}^{3'}\frac{\text{d}^3{\bf p}_{i}}{(2\pi)^32E_{i}} \frac{\text{d}^3{\bf p}_{\pi}}{(2\pi)^32E_{\pi}} \frac{E_d\text{d}^3{\bf r}}{m_d}  \notag \\
&&\times    {2m_dW_d(\tilde{\bf r},\tilde{\bf p})\big{(}\overline{|\mathcal{M}_{\pi^+ n\rightarrow \pi^+ n}|^2} + n\leftrightarrow p\big{)}}  \notag\\
&&\times \Big{[} -\big{(}\prod_{i=1'}^{3'} (1\pm f_i)\big{)}g_\pi f_\pi g_df_d +\frac{3}{4}\big{(}\prod_{i=1'}^{3'} g_if_{i}\big{)}  \notag\\
&&~~~~ \times (1+f_\pi)(1+f_d)\Big{]}\times  (2\pi)^4\delta^4(p_\text{in}-p_\text{out}), \label{eq:deu_rate}
\end{eqnarray}
from which ${\mathcal K}^>$ and ${\mathcal K}^<$ in Eq.~(\ref{eq:deu_kin}) can be identified. In the above equation, the $1\pm f_i$ in the square bracket is due to the quantum statistics of fermions ($-$) and bosons ($+$), and the $\delta$-function  denotes the conservation of energy and momentum with $p_\text{in}$ $=$ $\sum_{i=1'}^{3'}p_i$ and $p_{\rm out}$ $=$ $p_\pi +p_d$. The factor $3/4$ in the third  line is due to  the spin factors of initial and final states. The second line of Eq.~(\ref{eq:deu_rate}) denotes the spin-averaged squared amplitude between two scattering nucleons that are separated by the distance ${\bf r}$,  resulting in a nonlocal collision integral.  We note that the nonlocal collisions are important   for  studying spin polarizations  in nuclear collisions~\cite{Liu:2019krs,Weickgenannt:2020aaf}. The $W_d$ denotes the deuteron Wigner function, with $\tilde{\bf r}$ $=$ $\tilde{\bf r}_n-\tilde{\bf r}_p$ and $\tilde{\bf p}$ $=$ $(\tilde{\bf p}_n-\tilde{\bf p}_p)/2$ being, respectively, the relative coordinate and momentum in the  center-of-mass frame of the neutron and proton that forms the deuteron.  For simplicity, we take $W_d$ $=$ $8e^{-\tilde{\bf r}^2/\sigma_d^2-\tilde{\bf p}^2\sigma_d^2}$ with $\sigma_d$ $=$ $3.2~\rm fm$ to reproduce the empirical deuteron r.m.s. radius   of $r_d=1.96~\rm fm$~\cite{SchPRC59}.

For an approximately uniform system, the spatial part of $W_d$ in Eq.~(\ref{eq:deu_rate}) can be integrated out,   leading to $|\phi_d(\tilde{\bf p})|^2$ $=$ $\int \text{d}^3 {\bf r}  \gamma_d W_d$ $=$ $(4\pi\sigma_d^2)^{3/2}e^{-\tilde{\bf p}^2\sigma_d^2}$ with $\gamma_d$ $=$ $E_d/m_d$. This   reduces the second line of Eq.~(\ref{eq:deu_rate}) to $ 2m_d |\phi_d|^2\big{(}\overline{|\mathcal{M}_{\pi^+ p\rightarrow \pi^+ p}|^2}+ p\leftrightarrow n\big{)}$, which is the usual impulse approximation for $\overline{|\mathcal{M}_{\pi^+ d\rightarrow \pi^+ np}|^2}$, e.g. adopted in Ref.~\cite{DanNPA533}.  
 
The pion-nucleon scattering  matrix element  $\overline{|\mathcal{M}_{\pi^+ p\rightarrow \pi^+ p}|^2}$ can  be related to the $\pi N$ scattering cross section~\cite{Zhang:2017nck}. Under the impulse approximation, the  deuteron  dissociation cross section by a pion of momentum $p_{\rm lab}$ is approximately given by $\sigma_{\pi^+ d\rightarrow \pi^+ np} \approx \sigma_{\pi^+ n\rightarrow \pi^+ n} + \sigma_{\pi^+ p\rightarrow \pi^+ p}$, which holds for extremely large $p_\text{lab}$~\cite{DanNPA533,PDG2020}. At low  $p_\text{lab}$, e.g. $0.3~\rm GeV$, one can introduce  a  renormaliztion factor $F_d$~\cite{DanNPA533} such that $\sigma_{\pi^+ d\rightarrow \pi^+ np}$ $=$ $F_d(\sigma_{\pi^+ n\rightarrow \pi^+ n}+\sigma_{\pi^+ p\rightarrow \pi^+ p})$. 
As shown in the supplemental material~\cite{supp}, using the constant values $F_d$ $\approx$ $0.72$ and $F_{^3\text{He}}$ $\approx$ $0.51$ leads to an excellent description of the data for the $\pi{d}$ and $\pi^3\text{He}$ dissociation cross sections in the energy   regime  relevant for the present study. 

To numerically solve Eq.~(\ref{eq:deu_kin}) with the nonlocal collision integral given by Eq.~(\ref{eq:deu_rate}), we adopt the test particle ansatz ~\cite{WonPRC25}, i.e., mimicking the distribution functions $f_{\alpha}$ of a certain particle species of number $N_\alpha$ by a large number of delta functions, $f_{\alpha}({\bf r},{\bf p})$ $\approx$ $(2\pi)^3/(g_\alpha N_{\rm test})\sum^{N_{\alpha}N_\text{test}}_{i=1}\delta({\bf r}_i - {\bf r})\delta({\bf p}_i - {\bf p})$.
The $g_\alpha$ and $N_{\rm test}$ denote the spin degeneracy factor and number of test particles, respectively. The following stochastic method~\cite{DanNPA533,XZPRC71,WRFiP8} is then used to evaluate the collision integrals. To ensure the convergence of  numerical results, a sufficiently large $N_\text{test}$ will be used. 

Given the above rates for  deuteron dissociation and regeneration by a pion, the probability for the reaction $\pi^+d \rightarrow \pi^+np$ between a pion and a deuteron inside a volume $\Delta V$ to take place within a time interval $\Delta t$ can be obtained  as~\cite{DanNPA533,XZPRC71,WRFiP8} 
\begin{eqnarray}
P_{23}\big{|}_\text{IA} \approx F_d{ v}_{\pi^+ p}\sigma_{\pi^+ p\rightarrow \pi^+ p}\frac{\Delta t}{N_\text{test}{\Delta} V} + (p\leftrightarrow n)\label{eq:deu_p23},
\end{eqnarray}
where ${{v_{\pi^+ p}}}$ is the relative velocity between the pion and the proton inside the deuteron, and the two terms on the r.h.s correspond, respectively, to the two diagrams in Fig.~\ref{pic:piond}. Similarly, the probability for the reaction  $\pi^+np \rightarrow \pi^+d$ is 
\begin{eqnarray}
P_{32}\big{|}_\text{IA} \approx \frac{3}{4}F_d{ v}_{\pi^+p}\sigma_{\pi^+ p\rightarrow \pi^+ p}\frac{\Delta t W_d}{N_\text{test}^2\Delta V} +(p\leftrightarrow n)\label{eq:deu_p32_wig}.  
\end{eqnarray} 
Note that the deuteron Wigner function $W_d$ in Eq.~(\ref{eq:deu_p32_wig}) encodes the 
nonlocality of the scattering as it depends on both the coordinates and momenta of the constituent nucleons. It can be replaced by $|\phi_d|^2/(\gamma_d \Delta V)$ if  the hadronic matter size is much larger than the  deuteron size.

The above treatment for deuteron production and dissociation  can be   straightforwardly  generalized to those for $^3\text{H}$  from the   reactions $\pi NNN\leftrightarrow \pi ^3\text{H}$ and $\pi Nd\leftrightarrow \pi ^3\text{H}$, and similar ones for $^3\text{He}$. For the $3\leftrightarrow 2$ reaction, it can be similarly treated as for deuteron production.   The probability for the $4\to 2$ reaction  to occur in a volume $\Delta V$  within a time interval $\Delta t$ is, however, given by 
\begin{eqnarray}
P_{42}\big{|}_\text{IA}&\approx& \frac{1}{4}F_{^3\text{H}}\frac{{v}_{\pi N}\sigma_{\pi N \to\pi N}\Delta t}{N_\text{test}^3\Delta V} W_{^3\text{H}}, \label{eq:tri_p42}
\end{eqnarray}
where $F_{^3\text{H}}\approx F_{^3\text{He}} \approx 0.51$  and the triton Wigner function $W_{^3\text{H}}$ is also taken to  have  Gaussian forms as in Refs.~\cite{SchPRC59,CLWNPA729}. The branching ratio for the dissociation of $^3$H and $^3$He via the two inverse reactions is given in the supplemental  material~\cite{supp}. The above stochastic method reproduces well the thermally equilibrated deuteron and triton abundances in a box calculation with periodic boundary conditions~\cite{supp}. 

We note that there have been criticisms on the inclusion of light nuclei regeneration/dissociation reactions during the hadronic evolution, based on the naive argument that they have long formation times as a result of their small binding energies. This argument is, however, misleading as it is inconsistent with the time evolution of a many-nucleon system, which, according to Ref.~\cite{Rais:2022gfg}, shows that the light-nuclei components of its state vector change immediately once the interactions responsible for their formation are turned on.

\emph{Hadronic effects on light nuclei production in relativistic heavy-ion collisions.}{\bf ---}We first apply the above   kinetic  approach to central Au+Au collisions at $\sqrt{s_{\rm NN}}=200~\rm GeV$.  For the evolution of the QGP produced in these collisions, we use the collision-geometric-based 3D initial conditions \cite{Shen:2020jwv} with viscous hydrodynamic model MUSIC~\cite{PaqPRC93,SCPRC97,SCNST31} and a crossover type of equation of state at finite density NEOS-BQS~\cite{Monnai:2019hkn}. 
At hadronization of the QGP, both  hadrons  and light nuclei are produced using the SHM~\cite{AndNt561} with their phase-space distributions sampled on a constant energy density particlization hypersurface  according to the Cooper-Frye formula~\cite{CooPRD10}.    For the evolution of the hadronic matter, besides the nonlocal many-body scatterings for deuteron and triton regeneration and dissociation, most hadronic scattering channels in  the  ART model of the AMPT model~\cite{LZWPRC72} have been included. 

\begin{figure}[!t]
  \centering 
 \includegraphics[width=6.1cm]{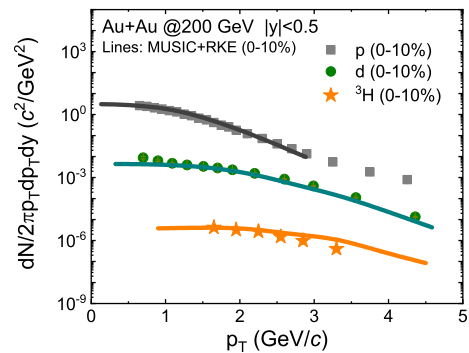}
  \caption{Transverse momentum spectra of $p$, $d$, and $^3\text{H}$  in central  Au+Au collisions at $\sqrt{s_{\rm NN}}=200~\rm GeV$. The experimental data, denoted by filled symbols,  are taken from Refs.~\cite{STARPRL97,STARPRC99,Zhang:2020ewj}, while 
 theoretical results  from the present kinetic approach (MUSIC+RKE) are  shown   by colored lines.
  }
  \label{pic:auau_sp}
\end{figure}

\begin{figure}[!t]
  \centering 
 \includegraphics[width=7.0cm]{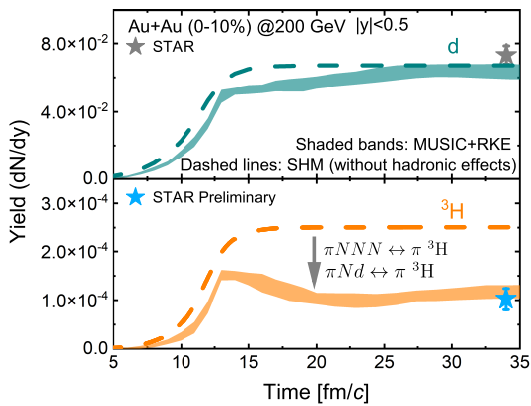}
  \caption{Time dependence of the yields of deuteron (upper panel) and triton (lower panel) in central Au+Au collisions at $\sqrt{s_{NN}}=200$ GeV. Results from the SHM at chemical freezeout and the present kinetic approach are  shown by dashed lines and shaded bands, respectively. The  experimental  data, denoted by filled stars, are from Refs.~\cite{STARPRC99,Zhang:2020ewj}.
  }
  \label{pic:auau_yield}
\end{figure}

Figure~\ref{pic:auau_sp} shows the comparison of  the transverse momentum ($p_T)$ spectra of $p$, $d$, and $^3\text{H}$ between the results of our kinetic approach (MUSIC+RKE) and the experimental data~\cite{STARPRL97,STARPRC99,Zhang:2020ewj} in central Au+Au collisions at $\sqrt{s_{\rm NN}} = 200~\rm GeV$, and the two are seen to agree very well. Figure~\ref{pic:auau_yield} further shows the time evolution of the yields of deuteron (upper panel) and triton (lower panel) from MUSIC+RKE (shaded bands) and its comparison with the (preliminary) experimental data from the STAR Collaboration~\cite{STARPRC99,Zhang:2020ewj}. The dashed lines denote the initial deuteron and triton yields predicted by the SHM. In these collisions, the final deuteron yield is about $95\%$ of its initial value given by the SHM, which confirms the results in a recent transport model study of deuteron production~\cite{OliPRC99}.  The small hadronic effect on the deuteron abundance is due to similar deuteron dissociation and regeneration rates during the hadronic evolution. The hadronic effects are, however, no longer small for the final triton yield. With the inclusion of $\pi NNN\leftrightarrow \pi^3\text{H}$ and $\pi Nd\leftrightarrow \pi ^3\text{H}$ reactions, the final triton yield from our kinetic approach is about half of its initial value at hadronization, which agrees  well with the experimental data. A similar result is obtained for helium-3.

The decreased $^3\text{H}$ and $^3\text{He}$ yields during the hadron evolution can be understood as follows. Due to the large dissociation and regeneration rates, deuterons and tritons are expected to  remain in chemical equilibrium with nucleons during the early stage of hadronic matter expansion. During this stage, the equilibrated triton number ($N_{^3\text{H}}$) can be related to the proton number ($N_p$)  and deuteron number ($N_d$) by $N_{^3\text{H}}\approx \alpha N_d^2/N_p$ where the constant $\alpha\approx$ 0.29 and
4/9 for a uniform~\cite{Sun:2020pjz,Shuryak:2020yrs} and large Gaussian source~\cite{Zhao:2021dka,Wu:2022cbh}, respectively.    With the deuteron number in our kinetic approach and also in Ref.~\cite{OliPRC99} remaining unchanged during the hadronic evolution, the triton number then has to decrease because of the more than a factor of 2 increase in the proton number from resonance decays.  
These results are consistent with those expected from an expanded hadronic matter with increasing total entropy but  constant baryon entropy~\cite{Xu:2017akx}, which would result in an increasing proton entropy due to resonance decays and lead to a decreasing $N_d/N_p$ ratio~\cite{Siemens:1979dz} and thus a stronger decrease in the $N_{^3\text{H}}/N_p$ ratio.
 Our results on decreased $^3\text{H}$ and $^3\text{He}$ yields are different, however, from the findings in a recent work using the rate equation~\cite{Neidig:2021bal} in a simple isentropic expansion model for the  hadronic matter, which  shows  that the triton and helium-3 yields remain almost unchanged during the hadronic evolution.

\begin{figure}[!t]
  \centering 
 \includegraphics[width=7.3cm]{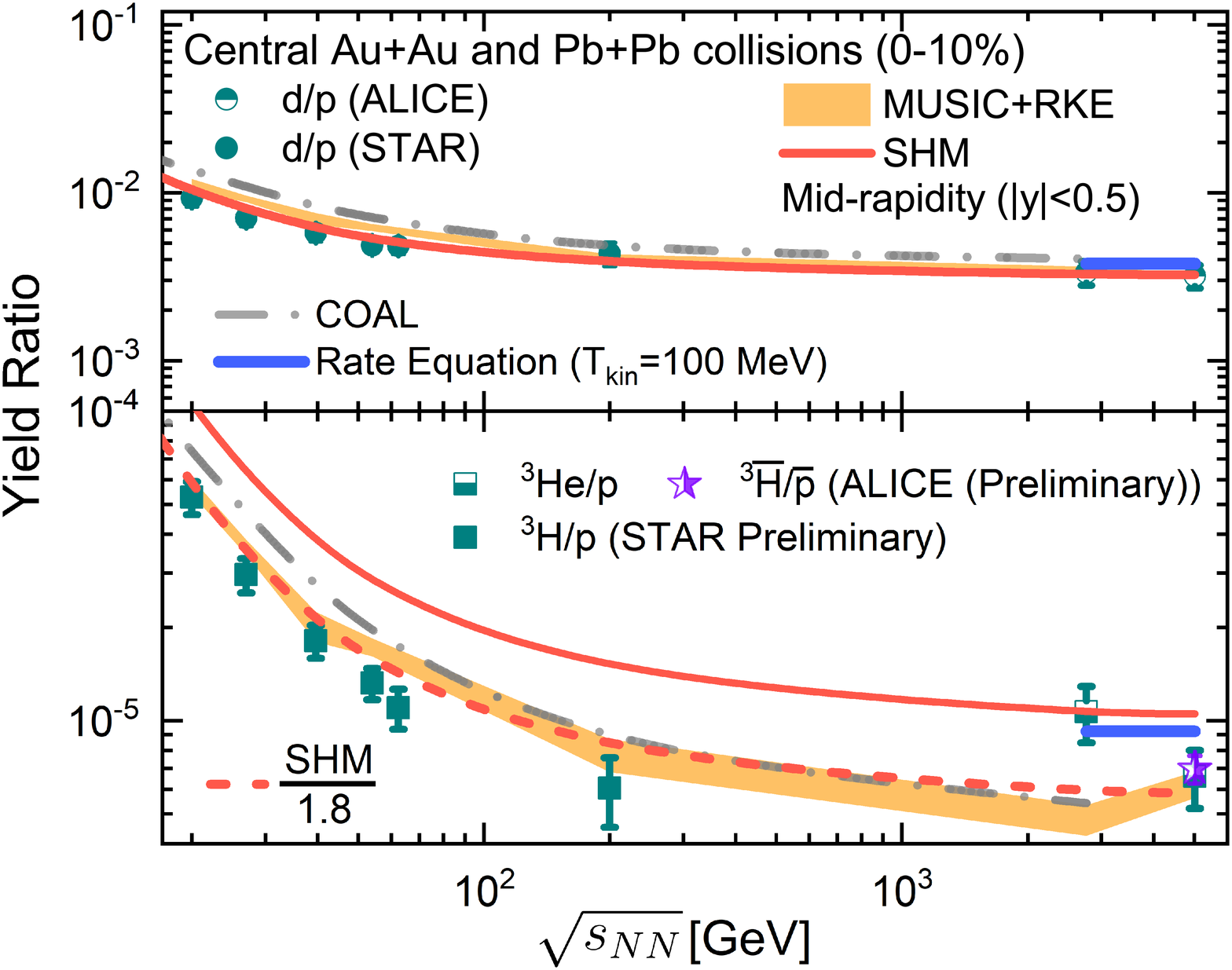}
  \caption{Collision energy dependence of light nuclei yield ratios. The shaded bands are results from the  present kinetic approach (MUSIC+RKE). Results from the coalescence model ~\cite{Zhao:2021dka} and the rate equation ~\cite{Neidig:2021bal} are denoted by the solid and dashed lines, respectively.  Experimental data denoted by symbols are from the STAR Collaboration~\cite{Zhang:2020ewj} and the ALICE Collaboration~\cite{ALICE:2015wav,Bartsch:2020lds,Bartsch:2022zyi}.  }
  \label{pic:ratio}
\end{figure}

To study the collision energy dependence of the hadronic effects, we now extend the above calculation  to central Au+Au and Pb+Pb collisions at $\sqrt{s_{NN}}=20-5020$ GeV. Figure~\ref{pic:ratio} shows the energy dependence of the light nuclei yield ratios  $d/p$, $^3\text{H}$/$p$, and  $^3\text{He}$/$p$. The shaded bands are results from  MUSIC+RKE. The (preliminary) experimental data, denoted by symbols, are taken from the STAR Collaboration~\cite{Zhang:2020ewj} and the ALICE Collaboration~\cite{ALICE:2015wav,Bartsch:2020lds,Bartsch:2022zyi}. As to deuteron production, results from both the SHM and our kinetic approach agree well with the experimental data within uncertainties  for all collision energies. For  triton or helium-3 production, results from the SHM describe well the experimental data point at $\sqrt{s_{NN}}=2.76$ TeV but significantly overestimate all other data points. Similar results are obtained from the rate equation~\cite{Neidig:2021bal} to account for the hadronic effects with a kinetic freezeout temperature of $T_\text{kin}\approx 100$ MeV.  On the other hand, results from our kinetic approach are about a factor of 1.8  (red dashed line) smaller than the SHM predictions and can nicely describe the experimental data except for the single data point from central Pb+Pb collisions at $\sqrt{s_{NN}}=2.76~$TeV.

We further show in Fig.~\ref{pic:ratio}  by dash-dotted lines the results from the nucleon coalescence model (COAL)~\cite{Zhao:2021dka}, in which light nuclei are produced at the kinetic freezeout  from the coalescence of two or three nucleons nearby in phase space~\cite{SchPRC59,Hillmann:2021zgj}. It is seen that the deuteron and triton yields from the coalescence model are very close to those from our kinetic approach calculation.  This result can be understood from the deep connection between the coalescence model and the kinetic approach, as pointed out in Ref.~\cite{SchPRC59}. Considering the limit that the regeneration and dissociation rates are equal in the kinetic equation (Eq.~(\ref{eq:deu_rate})), one obtains the equilibrated deuteron abundance as~\cite{supp}
\begin{eqnarray}
N_d \approx \frac{3}{4}\int d\Gamma_{np} g_nf_n({\bf r}_n,{\bf p^*}_n)g_pf_p({\bf r}_p,{\bf p}_p)W_d(\tilde{\bf r},\tilde{\bf p}), \label{eq:deu_coallimit}
\end{eqnarray}
where $d\Gamma_{np}$ $=$ $(2\pi)^{-6}\text{d}^3{\bf r}_n\text{d}^3{\bf p}_n\text{d}^3{\bf r}_p\text{d}^3{\bf p}_p$ and the neutron is chosen to be off mass shell to conserve energy.  Eq.~(\ref{eq:deu_coallimit}) is seen to resemble the deuteron yield in  the phase-space coalescence model~\cite{SchPRC59}. It should be mentioned, however, that the two nucleons forming the deuteron in the present kinetic calculation can be $nn$, $np$,  and $pp$ pairs, while only the $np$ pair is considered in the coalescence model~\cite{Zhao:2021dka}.

\emph{Summary.}{\bf ---} 
We have developed a  comprehensive kinetic approach to describe the dynamics of nucleosynthesis in relativistic heavy-ion collisions  by including the finite sizes of nuclei and using a stochastic method to evaluate the nonlocal many-body collision integrals in the relativistic kinetic equations. Our kinetic approach is seen to  describe well  deuteron and triton production in central Au+Au and Pb+Pb collisions at $\sqrt{s_{NN}}=20-5020$ GeV.  Most importantly, we have found that including light nuclei regeneration/dissociation reactions ($\pi NN\leftrightarrow\pi d$ and $\pi NNN\leftrightarrow\pi ^3\text{H}~(^3\text{He})$) during the hadronic matter expansion reduces the triton and helium-3 yields by   a factor of 1.8 from their initial values given by the   statistical hadronization model.   This finding naturally resolves the overestimation of triton production in the statistical hadronization model and provides thus the  evidence  for  ``dynamics at work"~\cite{Schukraft:2017nbn} in the nucleosynthesis   during  relativistic heavy-ion collisions. Our model calculations have suggested that the light nuclei yields are determined at the late stage of hadronic evolution, which justifies the assumption of the nucleon coalescence model. Extensions of the present study to (anti-)helium-4 as well as other exotic QCD molecular states like (anti-)hypertriton~\cite{ALargeIonColliderExperiment:2021puh} and X(3872)~\cite{ExHIC:2010gcb,Wu:2020zbx,Zhang:2020dwn} are of particular interest for future studies.

\begin{acknowledgments} 
We thank Lie-Wen Chen, Zi-Wei Lin, Xiaofeng Luo, Feng Li,  Zhen Zhang, Xiao-Jian Du, Shuai Liu, Shanshan Cao, and Wenbin Zhao for helpful discussions, and Chen Zhong for setting up and maintaining the GPU server. This work was supported in part by the U.S. Department of Energy under Award No. DE-SC0015266, No. DE-SC0013460, the Welch Foundation under Grant No. A-1358, the National Science Foundation under grant number PHY-2012922, National Natural Science Foundation of China under contract No. 11891070, No. 11890714 and No. 12147101.
\end{acknowledgments}


\end{document}